\documentstyle[12pt]{article}
\begin{document}

\def\ds{\displaystyle}
\def\beq{\begin{equation}}
\def\eeq{\end{equation}}
\def\bea{\begin{eqnarray}}
\def\eea{\end{eqnarray}}
\def\beeq{\begin{eqnarray}}
\def\eeeq{\end{eqnarray}}
\def\ve{\vert}
\def\vel{\left|}
\def\ver{\right|}
\def\nnb{\nonumber}
\def\ga{\left(}
\def\dr{\right)}
\def\aga{\left\{}
\def\adr{\right\}}
\def\lla{\left<}
\def\rra{\right>}
\def\rar{\rightarrow}
\def\nnb{\nonumber}
\def\la{\langle}
\def\ra{\rangle}
\def\ba{\begin{array}}
\def\ea{\end{array}}
\def\tr{\mbox{Tr}}
\def\ssp{{\Sigma^{*+}}}
\def\sso{{\Sigma^{*0}}}
\def\ssm{{\Sigma^{*-}}}
\def\xis0{{\Xi^{*0}}}
\def\xism{{\Xi^{*-}}}
\def\qs{\la \bar s s \ra}
\def\qu{\la \bar u u \ra}
\def\qd{\la \bar d d \ra}
\def\qq{\la \bar q q \ra}
\def\gGgG{\la g^2 G^2 \ra}
\def\q{\gamma_5 \not\!q}
\def\x{\gamma_5 \not\!x}
\def\g5{\gamma_5}
\def\sb{S_Q^{cf}}
\def\sd{S_d^{be}}
\def\su{S_u^{ad}}
\def\ss{S_s^{??}}
\def\sbp{{S}_Q^{'cf}}
\def\sdp{{S}_d^{'be}}
\def\sup{{S}_u^{'ad}}
\def\ssp{{S}_s^{'??}}
\def\sig{\sigma_{\mu \nu} \gamma_5 p^\mu q^\nu}
\def\fo{f_0(\frac{s_0}{M^2})}
\def\ffi{f_1(\frac{s_0}{M^2})}
\def\fii{f_2(\frac{s_0}{M^2})}
\def\O{{\cal O}}
\def\sl{{\Sigma^0 \Lambda}}
\def\es{\!\!\! &=& \!\!\!}
\def\ap{\!\!\! &\approx& \!\!\!}
\def\ar{&+& \!\!\!}
\def\ek{&-& \!\!\!}
\def\kek{\!\!\!&-& \!\!\!}
\def\cp{&\times& \!\!\!}
\def\se{\!\!\! &\simeq& \!\!\!}
\def\eqv{&\equiv& \!\!\!}
\def\kpm{&\pm& \!\!\!}
\def\kmp{&\mp& \!\!\!}

% .........................................................

\def\simlt{\stackrel{<}{{}_\sim}}
\def\simgt{\stackrel{>}{{}_\sim}}

% .........................................................

\title{
         {\Large
                 {\bf
Semileptonic $D_s^+ \rar \phi \bar{\ell} \nu$ decay in 
QCD light cone sum rule 
                 }
         }
      }

\author{\vspace{1cm}\\
{\small T. M. Aliev$^a$ \thanks
{e-mail: taliev@metu.edu.tr}\,\,,
A. \"{O}zpineci$^b$ \thanks
{e-mail: ozpineci@ictp.trieste.it}\,\,,
M. Savc{\i}$^a$ \thanks
{e-mail: savci@metu.edu.tr}} \\
{\small a Physics Department, Middle East Technical University,
06531 Ankara, Turkey}\\
{\small b  The Abdus Salam International Center for Theoretical Physics,
I-34100, Trieste, Italy} }
\date{}

\begin{titlepage}
\maketitle
\thispagestyle{empty}

\begin{abstract}
We calculate the form factors $V$, $A_1$, $A_2$ and $A_0$ appearing in  
the $D \rar \phi$ transition in light cone QCD sum rule method. We compare
our results on these form factors with the current experimental results and
existing theoretical calculations.
\end{abstract}

%\vspace{1cm}
~~~PACS numbers: 11.55.Hx, 13.20.Fc
\end{titlepage}

\section{Introduction}

Semileptonic decays of mesons containing charm and beauty quarks constitute
a very important class of decays for studying strong and weak interactions.
These decay modes of heavy flavored mesons are much more clear samples
compared to that of the hadronic decay modes, since leptons do not
participate in the strong interaction.

Therefore, the study of these decays is one efficient way for
determining the elements of the Cabibbo--Kobayashi--Maskawa (CKM) matrix,
as well as for understanding the origin of CP violation which is related to
the structure of the CKM matrix in Standard Model (SM). 

An accurate determination of CKM matrix elements, obviously, depends
crucially on the possibility of controlling the effects of the strong
interactions. For exclusive decays, where initial and final states
of hadrons are known, the main job is to calculate various transition form
factors, which involve all the long distance QCD dynamics. So, some
non--perturbative approach for estimating the long distance effects is
needed. Several methods have been used to treat these effects, such as quark
model, QCD sum rules, lattice theory, chiral perturbation theory, etc. Among
these approaches, QCD sum rules occupies a special place, since it is based
on the very first principles of QCD. 

The method of QCD sum rules \cite{R6201} has been successfully applied to
wide variety of problems of hadron physics (see \cite{R6202,R6203} and
references therein). In this method, physical observables of hadrons are
related to QCD vacuum via a few condensates. The semileptonic decay 
$D \rar \bar{K}^0 e \bar{\nu}_e$ was firstly studied in QCD sum rules with
3--point correlation function in \cite{R6204}. This method, then,
successfully extended to study other semileptonic decay decays of $D$ and $B$
mesons, i.e., $D^+ \rar \bar{K}^0 e^+ \nu_e$, $D^+ \rar \bar{K}^{0\ast} e^+ \nu_e$
\cite{R6205}, $D \rar \pi e \bar{\nu}_e$, $D \rar \rho e \bar{\nu}_e$
\cite{R6206}, $B \rar D(D^\ast) \ell \bar{\nu}_\ell$ \cite{R6207} and
$B \rar \pi \ell \bar{\nu}_\ell$ \cite{R6208}.

However, this method inherits some problems, the main one being 
that some of the form factors have nasty behavior in the heavy quark limit
$m_Q \rar \infty$. In order to overcome the problems of the traditional QCD
sum rules, an alternative method, namely light cone QCD sum rules (LCQSR)
was developed in \cite{R6209} and is regarded as an efficient tool in
studying exclusive processes which involve the emission of a light particle.

The LCQSR is based on the operator product product expansion (OPE) near the
light cone $x^2 \approx 0$, which is an expansion over the twist of the
operators, rather than the dimensions as in the traditional QCD sum rules.
All non--perturbative dynamics is parametrized by the so--called light cone
wave functions, instead of the vacuum condensates in the traditional sum
rules, which represents the matrix elements of the nonlocal operators
between the vacuum and the corresponding particle (more about this method
can be found in \cite{R6203,R6210})

The LCQSR has wide range of applications to numerous problems of hadron
physics. One of the promising ways for obtaining information about CKM
matrix elements, as well as about wave functions, is studying the
semileptonic decays. 

In this work we study $D_s^+ \rar \phi \bar{\ell} \nu$ decay in LCQSR. This
decay mode has been measured in experiments in \cite{R6211}--\cite{R6214}.
Note that $D \rar \phi$ transition form factors are  calculated in the 
framework of traditional 3--point QCD sum rules in \cite{R6215,R6216}, but
the results don't confirm each other. Therefore we decided to calculate 
$D \rar \phi$ form factors using light cone sum rules as an alternative 
approach to the traditional sum rules.

The paper is organized as follows. In section 2 we derive the sum rules for
the transition form factor. Section 3 is devoted to the numerical analysis
and discussions and contains a summary of results and conclusions.

\section{Light cone sum rules for the $D_s \rar \phi$ transition form
factors}

We start by defining the form factors of  $D_s \rar \phi$ weak form factors
in the following way

\bea
\label{e6201}
\lefteqn{
\la \phi(P) \ve \bar{s} \gamma_\mu (1-\gamma_5) c \ve D_s(p_{D_s}) \ra =
- i \varepsilon_\mu^\ast (m_{D_s} + m_\phi) A_1(q^2) } \nnb \\
\ar i (p_{D_s} + P)_\mu (\varepsilon^\ast q) 
\frac{A_2(q^2)}{m_{D_s} + m_\phi} + i q_\mu (\varepsilon^\ast q) 
\frac{2 m_\phi}{q^2} [A_3(q^2)-A_0(q^2)] \nnb \\
\ar \frac{2 V(q^2)}{m_{D_s} + m_\phi} \epsilon_{\mu\alpha\beta\gamma} 
\varepsilon^{\ast\alpha} q^\beta P^\gamma~,
\eea
where $q=p_{D_s}-P$ is the momentum transfer, $P$ and $\varepsilon$ are
four momentum vector polarization of the vector $\phi$ meson, respectively,
and $p_{D_s}$ is the four momentum of $D_s$ meson.

In this section we derive sum rules for these form factors. In order to
calculate the form factors of the semileptonic $D_s \rar \phi \ell \nu$
decay, we consider the following correlator function
\bea
\label{e6202}
\Pi_\mu(P,q) \es i \int d^4x e^{iqx} \la \phi(P) T \big[\bar{s}(x) \gamma_\mu
(1-\gamma_5) c(x) \bar{c}(0) (1-\gamma_5) s(0) \big]\ve 0 \ra \nnb \\
\es \Gamma^0 \varepsilon_\mu^\ast - \Gamma^+ \frac{\varepsilon^\ast q}{Pq}
(2 P + q)_\mu - \Gamma^- \frac{\varepsilon^\ast q}{Pq} q_\mu +
i \Gamma^V \varepsilon_{\mu\alpha\beta\gamma} \varepsilon^{\ast \alpha}
q^\beta P^\gamma~.
\eea
The Lorentz invariant functions $\Gamma^{0,\pm,V}$ can be calculated in QCD
for large Euclidean $p_{D_s}^2$, to put it more correctly, when 
$m_c^2-p_{D_s}^2 \ll 0$, the correlation function (\ref{e6201}) is dominated
by the region of small $x^2$ and can be systematically expanded in powers of
deviation from the light cone $x^2=0$.

The main reason for choosing the chiral current $\bar{c} (1-\gamma_5) s$ is
that, in this case many of the twist--3 wave functions which are poorly
known and cause the main uncertainties to the sum rules, can effectively be
eliminated and provide results with less uncertainties. The chiral current
approach has been applied to studying $B \rar \pi$ \cite{R6217,R6218},
$B \rar \eta$ \cite{R6219} weak form factors.

Let us discuss firstly the hadronic representation of the correlator. This
can be done by inserting the complete set of intermediate states with the
same quantum numbers of the current operator $\bar{c} (1-\gamma_5) s$ in the
correlation function. By isolating the pole term of the lowest pseudoscalar
$D_s$ meson, we get the following representation of the correlator function
from hadron side
\bea
\label{e6203}
\Pi_\mu(P,q) \es \frac{\ds \la \phi \ve \bar{s} \gamma_\mu (1-\gamma_5) c
\ve D_s \ra \la D_s \ve \bar{c} (1-\gamma_5) s \ve 0 \ra}
{m_{D_s}^2 -(P+q)^2} \nnb \\
\ar \sum_h \frac{\ds \la \phi \ve \bar{s} \gamma_\mu (1-\gamma_5) c   
\ve h \ra \la h \ve \bar{c} (1-\gamma_5) s \ve 0 \ra}
{m_{h}^2 -(P+q)^2}~.
\eea
For the invariant amplitudes $\Gamma^{0,\pm,V}$, one can write a general
dispersion relation in the $p_{D_s}^2=(P+q)^2$ limit
\bea
\Gamma^i(q^2,(P+q)^2) = \int ds \frac{\rho^i(s)}{s-(P+q)^2} 
+ \mbox{\rm subtr.}~,\nnb
\eea
where the spectral densities corresponding Eq. (\ref{e6202}) can easily be
calculated. As an illustration of this fact, we present the result for
$\Gamma^0$
\bea
\label{e6204}
\rho^{(0)}(s) = \frac{f_{D_s} m_{D_s}^2}{m_c+m_s} (m_B + m_V) A_1(q^2)
\delta (s-m_{D_s}^2) + \rho^{(0)h}(s)~.
\eea
The first term in Eq. (\ref{e6204}) represents the contribution of the 
ground state $D_s$ meson. In deriving Eq. (\ref{e6202}), we have used
\bea
\la D_s \ve \bar{c} (1-\gamma_5) s \ve 0 \ra = i \frac{f_{D_s}
m_{D_s}^2}{m_c+m_s}~. \nnb
\eea
The second term in Eq. (\ref{e6204}) corresponds to the spectral density of
the higher resonances and continuum. The spectral density $\rho^{(0)h}(s)$
can be approximated by invoking the quark hadron duality anzats
\bea
\rho^{(0)h}(s) = \rho^{(0)QCD} (s-s_0)~. \nnb
\eea
So for the hadronic representation of the invariant amplitude $\Gamma^{(0)}$
we have
\bea
\label{e6205}
\Gamma^{(0)} = \frac{f_{D_s} m_{D_s}^2}{m_c+m_s} \frac{m_B + m_\phi}
{m_{D_s}^2-(P+q)^2} A_1(q^2) + \int_{s_0}^{\infty} ds
\frac{\rho^{(0)QCD}(s)}{s-(P+q)^2} + \mbox{\rm subtr.}~.
\eea
Hadronic representations for other invariant amplitudes can be constructed
in precisely the same manner.

In order to obtain sum rules for the form factors $A_1$, $A_2$, $A_0$ and
$V$, we must calculate the correlator from QCD side. This calculation can be
performed by using the light cone OPE. The contributions to OPE can be
obtained by contracting the quark fields to a full c--quark propagator,
i.e.,
\bea
\label{e6206}
\Pi_\mu (P,q) \es i \int d^4x e^{iqx} \la \phi \ve \bar{s} \gamma_\mu
(1-\gamma_5) S_c(x) (1-\gamma_5) s(0) \ve 0 \ra \nnb \\
\es \frac{i}{4} \int d^4x e^{iqx} \Big[\mbox{\rm Tr} \gamma_\mu (1-\gamma_5)
S_c(x)(1-\gamma_5) \Gamma_i \Big]\la \phi \ve \bar{s} \Gamma^i s \ve 0 \ra~, 
\eea
where $\Gamma^i$ is the full set of the Dirac matrices $\Gamma^i =
(I,~\gamma_5,~\gamma_\alpha,~\gamma_\alpha
\gamma_5,~\sigma_{\alpha\beta})$, and 
\bea
\label{e6207}
i S_c(x) \es i S_c^{(0)}(x) - ig_s \int \frac{d^4k}{(2\pi)^4} e^{-ikx} 
\int_0^1 du \, \Bigg[ \frac{1}{2} \frac{\not\!k + m_c}{(m_c^2-k^2)^2}
G_{\mu\nu} (ux) \sigma^{\mu\nu} \nnb \\
\ar \frac{1}{m_c^2-k^2} u x_\mu G^{\mu\nu} (ux) \gamma_\nu \Bigg]~.
\eea
Here, $G_{\mu\nu}$ is the gluonic field strength, $g_s$ is the strong
coupling constant and $S_c^{(0)}$ represents a free c--quark propagator
\bea
\label{e6208}
S_c^{(0)}(x) = \int \frac{d^4k}{(2\pi)^4} e^{-ikx} \frac{\not\!k +
m_c}{k^2-m_c^2}~.
\eea
From Eqs. (\ref{e6206})--(\ref{e6208}) we see that, in order to calculate
the theoretical part of the correlator, the matrix elements of the nonlocal
operators between vector $\phi$ meson and vacuum states are needed. We see
from Eq. (\ref{e6206}) that, the contribution to the correlator comes only
from the wave functions that contain odd--number $\gamma$ matrices. 

Up to twist--4, the $\phi$ meson wave functions containing odd--number of
$\gamma$ matrices, and appearing in our calculations are:
\bea
\label{e6209}
\la \phi(P,\lambda) \ve \bar{s}(x) \gamma_\mu s(0) \ve 0 \ra \es 
f_\phi m_\phi \Bigg[ P_\mu \frac{e^\lambda x}{Px} \int_0^1 du \,  
e^{iuPx} \Bigg( \Phi_\parallel(u,\mu^2) +
\frac{m_\phi^2 x^2}{16} A(u,\mu^2) \Bigg) \nnb \\ 
\ar \Bigg(e^\lambda_\mu-P_\mu \frac{e^\lambda x}{Px} \Bigg) \int_0^1 
du\, e^{iuPx} g_\perp^{(v)}(u,\mu^2) \nnb \\ 
\ek \frac{1}{2} x_\mu
\frac{e^\lambda x}{(Px)^2} m_\phi^2 \int_0^1 du\, e^{iuPx}
C(u,\mu^2)~, \\ \nnb \\
\label{e6210}
\la \phi(P,\lambda) \ve \bar{s}(x) \gamma_\mu \gamma_5 s(0) \ve 0 \ra \es
\frac{1}{4} \Bigg( f_\phi - \frac{2 f_\phi^T m_s}{m_\phi} \Bigg) m_\phi 
\epsilon_\mu^{\phantom{\mu}\nu\alpha\beta} e_\nu^\lambda P_\alpha x_\beta
\int_0^1 du \, e^{iuPx} g_\perp^{(a)}(u,\mu^2)~, \nnb \\ \\
\label{e6211}
\la \phi(P,\lambda) \ve \bar{s}(x) g G_{\mu\nu}(ux) i \gamma_\alpha s(0) \ve 0 \ra \es
f_\phi m_\phi p_\alpha (p_\nu e_{\perp \mu}^\lambda - 
p_\mu e_{\perp \nu}^\lambda ){\cal V} (u,px) \nnb \\
\ar f_\phi m_\phi^3 \frac{e^\lambda x}{px} (p_\mu g_{\alpha \nu}^\perp - 
p_\nu g_{\alpha \mu}^\perp )\Phi (u,px) \nnb \\
\ar f_\phi m_\phi^3 \frac{e^\lambda x}{(px)^2} p_\alpha (p_\mu x_\nu -   
p_\nu x_\mu )\Psi (u,px)~,\\ \nnb \\
\label{e6212}
\la \phi(P,\lambda) \ve \bar{s}(x) g \widetilde{G}_{\mu\nu}(ux) i 
\gamma_\alpha \gamma_5 s(0) \ve 0 \ra \es
f_\phi m_\phi p_\alpha (p_\nu e_{\perp \mu}^\lambda - 
p_\mu e_{\perp \nu}^\lambda ) \widetilde{\cal V}(u,px) \nnb \\
\ar f_\phi m_\phi^3 \frac{e^\lambda x}{px} (p_\mu g_{\alpha \nu}^\perp -   
p_\nu g_{\alpha \mu}^\perp )\widetilde{\Phi} (u,px)\nnb \\
\ar f_\phi m_\phi^3 \frac{e^\lambda x}{px} p_\alpha (p_\mu x_\nu -          
p_\nu x_\mu )\widetilde{\Psi}(u,px)~.
\eea
In all expressions, we have used 
\bea
\label{e6213}
p_\mu \es P_\mu - \frac{1}{2} x_\mu \frac{m_\phi^2}{px}~, \nnb \\ 
e_\mu^\lambda \es \frac{e^\lambda x}{px} \Bigg(p_\mu - \frac{m_\phi^2}{2(px)}
x_\mu \Bigg) + e_{\perp \mu}^\lambda~, \nnb \\
g_{\mu\nu}^\perp \es g_{\mu\nu} - \frac{1}{px} 
(p_\mu x_\nu + p_\nu x_\mu)~, 
\eea     
where $\Phi_\parallel$ is the leading twist--2 wave function, while
$g_\perp^{(v)}$, $g_\perp^{(a)}$, ${\cal V}$ are twist--3 and
all the remaining ones are twist--4 wave functions. The notation used
in Eqs. (\ref{e6211})--(\ref{e6214}) is the following
\bea
\label{e6214}
K(u,Px) = \int {\cal D} \alpha e^{iPx(\alpha_1 + u \alpha_3)}
K(\alpha)~,
\eea
where 
\bea
{\cal D} \alpha = d\alpha_1 d\alpha_2 d\alpha_3 \delta(1-\alpha_1-\alpha_2
-\alpha_3)~. \nnb
\eea
Inserting Eqs. (\ref{e6207}) and (\ref{e6208}) into Eq. (\ref{e6206}) and
using the definitions of $\phi$ meson wave functions, the invariant
structures $\Gamma^{0,\pm,V}$ take the following form
\bea
\label{e6215}
\Gamma_0 \es \int du \, \frac{2 f_\phi m_\phi m_c
g_\perp^{(v)}(u)}{\Delta_1} ~, \\ \nnb \\
\label{e6216} 
\Gamma^+ \es \int du \, \Bigg\{ \frac{f_\phi m_\phi m_c}{\Delta_1^3}
\Big[ m_c^2 m_\phi^2 A(u) + 4 (Pq) \Delta_1 
\Phi_\parallel^{(i)} (u) \Big] -
\frac{f_\phi m_\phi^3 m_c u C(u)}{\Delta_1^2} \nnb \\
\ek \int {\cal D} \alpha 
\, \frac{f_\phi m_\phi^3 m_c}{\Delta_2^2} \Bigg(2 \Phi - 2 \widetilde{\Phi}
+\Psi - \widetilde{\Psi} - \frac{{\cal V}}{2} + \frac{\widetilde{\cal V}}{2}
\Bigg) \Bigg\}~, \\ \nnb \\
\label{e6217}
\Gamma^- \es \int du \, \Bigg\{ \frac{-f_\phi m_\phi m_c}{\Delta_1^3}
\Big[ m_c^2 m_\phi^2 A(u) + 4 (Pq) \Delta_1 
\Phi_\parallel^{(i)} (u) \Big] -      
\frac{f_\phi m_\phi^3 m_c (2-u) C(u)}{\Delta_1^2} \nnb \\
\ar \int {\cal D} \alpha
\, \frac{f_\phi m_\phi^3 m_c}{\Delta_2^2} \Bigg(2 \Phi - 2
\widetilde{\Phi} +\Psi - \widetilde{\Psi}  - \frac{{\cal V}}{2} +
\frac{\widetilde{\cal V}}{2} \Bigg) \Bigg\}~, \\ \nnb \\
\label{e6218}
\Gamma^V \es \int du \,  \Bigg(1-\frac{2 f_\phi^T m_s}{f_\phi m_\phi}
\Bigg)  g_\perp^{(a)} \frac{f_\phi m_\phi m_c}{\Delta_1^2}~,
\eea
where 
\bea
\Phi_\parallel^{(i)} (u) \es \int_0^u dv \Big[\Phi_\parallel(v) - g_\perp^{(v)}
(v) \Big]~, \nnb \\
\Delta_1 \es m_c^2 - (q+Pu)^2~, \nnb \\
\Delta_2 \es  m_c^2 - [q+(\alpha_1+u\alpha_3)P]^2~, \nnb
\eea

Equating expressions of the invariant structures $\Gamma^{0,\pm,V}$ coming
from QCD and phenomenological parts of the correlation function and making
the Borel transformation with respect to $(P+q)^2$ in both parts, in order to
suppress the contributions of higher states and continuum and also to
eliminate the subtraction terms in the dispersion integral, we get the
following sum rules for the $D \rar \phi$ transition form factors: 
\bea
\label{e6219}
A_1(q^2) \es \frac{m_c+m_s}{f_{D_s} m_{D_s}^2} \, \frac{1}{m_{D_s}+m_\phi}
e^{m_{D_s}^2/M^2} \Bigg\{2 f_\phi m_\phi m_c \int_\delta^1 du \,
\frac{g_\perp^{(v)}(u)}{u} e^{-s(u)/M^2}  \Bigg\}~, \\ \nnb \\
\label{e6220}
A_2(q^2) \es \frac{(m_c+m_s)(m_{D_s}+m_\phi)}{f_{D_s} m_{D_s}^2}
\,\frac{2}{m_{D_s}^2-m_\phi^2-q^2}
e^{m_{D_s}^2/M^2} \nnb \\
\cp \Bigg\{f_\phi m_\phi m_c \Bigg[ \frac{1}{2} m_\phi^2 m_c^2
\int_\delta^1 \frac{1}{u} A(u) \frac{1}{2(M^2 u)^2} e^{-s(u)/M^2}
- \int_\delta^1 du \, \frac{1}{u^2} \Phi_\parallel^{(i)}(u) 
e^{-s(u)/M^2} \nnb \\
\ar \int_\delta^1 du \, \frac{1}{u} (m_c^2-m_\phi^2 u^2-q^2) 
\frac{\Phi_\parallel^{(i)}(u)}{u^2 M^2} e^{-s(u)/M^2}
-m_\phi^2 \int_\delta^1 du \, \frac{u C^{(i)}(u)}{u^2 M^2} e^{-s(u)/M^2}
\Bigg] \nnb \\
\ek f_\phi m_\phi^3 m_c \int_\delta^1 du \, {\cal D} \alpha \, \theta(s_0-s(k)) 
\frac{1}{k^2 M^2} \Bigg[2 \Phi(\alpha)-2\widetilde{\Phi}(\alpha)
+ \Psi(\alpha)-\widetilde{\Psi}(\alpha) \nnb \\
\ek \frac{{\cal V}}{2} + \frac{\widetilde{\cal V}}{2}
\Bigg]e^{-s(k)/M^2}\Bigg\}~.
\eea

The form factor $A_3(q^2)$ can be obtained from the exact result
\bea
\label{e6221}
A_3(q^2) =\frac{m_{D_s}+m_\phi}{2 m_\phi} A_1(q^2) -
\frac{m_{D_s}-m_\phi}{2 m_\phi} A_2(q^2)~,
\eea
and $A_0(q^2)$ can be calculated from the following sum rule
\bea
\label{e6222}
A_3(q^2)\kek A_0(q^2) = \frac{m_c+m_s}{f_{D_s} m_{D_s}^2}\,      
\frac{q^2}{2 m_\phi} \,\frac{1}{m_{D_s}^2-m_\phi^2-q^2}
e^{m_{D_s}^2/M^2} \nnb\\
\cp \Bigg\{f_\phi m_\phi m_c \Bigg[ -\frac{1}{4} m_\phi^2 m_c^2
\int_\delta^1 \frac{1}{u} A(u) \frac{1}{2(M^2 u)^2} e^{-s(u)/M^2}
+ \int_\delta^1 du \, \frac{1}{u^2} \Phi_\parallel^{(i)}(u) 
e^{-s(u)/M^2} \nnb \\
\ek \int_\delta^1 du \, \frac{1}{u} (m_c^2-m_\phi^2 u^2-q^2) 
\frac{\Phi_\parallel^{(i)}(u)}{u^2 M^2} e^{-s(u)/M^2}
-m_\phi^2 \int_\delta^1 du \, \frac{(2-u)C^{(i)}(u)}{u^2 M^2} e^{-s(u)/M^2}
\Bigg] \nnb \\
\ar f_\phi m_\phi^3 m_c \int_\delta^1 du \, {\cal D} \alpha \, \theta(s_0-s(k)) 
\frac{1}{k^2 M^2} \Bigg[2 \Phi(\alpha)-2\widetilde{\Phi}(\alpha)
+ \Psi(\alpha)-\widetilde{\Psi}(\alpha) \nnb \\
\ek \frac{{\cal V}}{2} + \frac{\widetilde{\cal V}}{2}
\Bigg] e^{-s(k)/M^2} \Bigg\}~, \\ \nnb \\
\label{e6223}
V(q^2) \es \frac{(m_{D_s}+m_\phi)(m_c+m_s)}{2 f_{D_s} m_{D_s}^2} 
e^{m_{D_s}^2/M^2} \nnb \\
\cp \Bigg\{ \Bigg(1-\frac{2m_s f_\phi^T}{f_\phi m_\phi}\Bigg) f_\phi m_\phi m_c
\int_\delta^1 du \, g_\perp^{(a)}(u) \frac{1}{u^2 M^2} e^{-s(u)/M^2}
\Bigg\}~,
\eea
where $M^2$ is the Borel parameter and 
\bea
s(t) \es \frac{m_c^2-q^2\bar{t}+m_\phi^2 t \bar{t}}{t}~,\nnb \\
t \es \left\{ \begin{array}{l}
u,~~\mbox{\rm or}~,\nnb\\ \\
k=\alpha_1+u\alpha_3~,\end{array} \right. \nnb \\
\bar{t} \es 1-t~, \nnb \\
\delta \es \frac{1}{2 m_\phi^2} \Big[(m_\phi^2+q^2-s_0) +
\sqrt{(s_0-m_\phi^2-q^2)^2-4 m_\phi^2(q^2-m_c^2)}\Big]~.\nnb
\eea
\section{Numerical analysis}

In this section we present our numerical calculation of the form factors
$A_1$, $A_2$, $A_0$ and $V$. As can easily be seen from the expressions of
these form factors, the main input parameters are the $\phi$ meson wave
functions, whose explicit forms are given in \cite{R6220} and we use them in
our study. The values of the other input parameters appearing in sum rules
for form factors are: $m_{D_s} = 1.9686~GeV$, $m_s = 0.14~GeV$,
$m_c=1.3~GeV$, $f_{D_s} = (0.214 \pm0.038)~GeV$ \cite{R6203},
$m_\phi=1.02~GeV$. The leptonic decay constant of $\phi$ meson, which is
$f_\phi=0.234~GeV$, is extracted from the experimental result of the $\phi
\rar \ell^+ \ell^-$ decay \cite{R6221}. The threshold
$s_0=(6.5 \pm 0.5)~GeV^2$ is determined from the analysis of the three point
function sum rules for $f_{D_s}$ (see for example \cite{R6203}).

With the above--mentioned input parameters, we now proceed by carrying out
our numerical analysis. The first step, according to sum rules philosophy,
is to look a working region of the auxiliary Borel parameter $M^2$, where
numerical results should be stable for a given threshold $s_0$. The lower
limit of $M^2$ is determined by the condition that the terms $M^{-2n}~(n>1)$
remains subdominant. The upper bound of $M^2$ is determined by requiring
that the continuum and higher state contributions constitute maximum 30\% of
the total result. Our numerical analysis shows that both requirements are
satisfied in the region $3~GeV^2 \le M^2 \le 4.5~GeV^2$. We should note that
LCQSR for the form factors are reliable at the region $q^2 \simlt
0.4~GeV^2$. Moreover, we analyse the $M^2$ dependencies of the form factors
$A_1$, $A_2$, $A_0$ and $V$ at $q^2=0~GeV^2$ and
$q^2=0.2~GeV^2$, for three different values of the continuum threshold,
namely, $s_0=6.0,~6.5$ and $s_0=7.0~GeV^2$. Our analysis shows that
the form factors are practically independent of the Borel mass
when $M^2$ varies between $3~GeV^2$ and $4~GeV^2$. Variation of the form
factors in relation to the continuum threshold is also very weak. The
results for all form factors change about $5\%$ at $q^2=0$. Our final
results for the form factors at $q^2=0$ and $s_0=6.5~GeV^2$, are
\bea
\label{e6224}
A_1(0) \es 0.65 \pm 0.15~, \nnb \\    
A_2(0) \es 0.85 \pm 0.20~, \nnb \\
A_0(0)=A_3(0) \es 0.56 \pm 0.1~, \nnb \\
V(0) \es 0.90 \pm 0.20~.
\eea

It should be noted that in the region $q^2 \ge 0.4~GeV^2$ the applicability
of the light cone QCD sum rule is questionable. In order to extend our
results to the full physical region, we look for a parametrization of the
form factors in such a way that in the region $0 \le q^2 \le 0.4~GeV^2$,
the above--mentioned parametrization coincides with the light cone QCD sum
rules prediction. The most convenient parametrization of the $q^2$ dependence 
of the form factors is given in terms of three parameters in the following form
\bea
\label{e6225}
F_i(q^2) = \frac{\ds F_i(0)}{\ds 1-a_{F_i} (q^2/m_{D_s}^2) + 
b_{F_i} (q^2/m_{D_s}^2)^2}~.
\eea
The values of the parameters $F_i(0)$, $a_{F_i}$ and $b_{F_i}$ are listed in
Table (1).
 
\begin{table}[h]     
\renewcommand{\arraystretch}{1.5}
\addtolength{\arraycolsep}{3pt}
$$
\begin{array}{|l|ccc|}
\hline
& F(0) & a_F & b_F \\ \hline
A_1 &
0.65 & 1.36 & -0.31 \\
A_2 &
0.85 & 4.5 &\phantom{-}5.55 \\
A_0 &
0.56 & 0.13 & -0.46 \\
V &
0.9 & 2.82 &\phantom{-}1.51 \\ \hline
\end{array}
$$
\caption{Parameters of the form factors given in Eq. (\ref{e6225}), for the
$D_s$ decay in a three--parameter fit. We take the central values of the
form factors for $F(0)$.}
\renewcommand{\arraystretch}{1}
\addtolength{\arraycolsep}{-3pt}
\end{table}
We proceed by discussing the uncertainties related to the input parameters
and wave functions. We note first that the radiative corrections to the
leading twist--2 function, which is calculated in \cite{R6220}, is about
$\sim 10\%$. As has already been noted, the results depend weakly on the
continuum threshold $s_0$ and Borel parameter $M^2$, and the uncertainty due
to these parameters is about $5\%$--$7\%$ in the working region of $M^2$.
Moreover, the results are also quite weakly dependent on the vector meson
decay constant $f_\phi$ and $f_\phi^T$, which results in an uncertainty
about $5\%$. Additional uncertainty coming from the Gegenbauer moments are
about $\sim 10\%$. Summing up all these above--mentioned errors, the overall
uncertainty in the values of the form factors is of the order of $17\%$.

In the experiments, the ratios
\bea
r_1 = \frac{V(0)}{A_1(0)}~,~~\mbox{\rm and}~~~
r_2 = \frac{A_2(0)}{A_1(0)}~, \nnb
\eea
are measured. In the present work, within the framework of the light cone
QCD sum rules, we get $r_1 = 1.57 \pm 0.36$ and 
$r_2 = 1.30 \pm 0.34$. In Table (2), we present a comparison of our
results with the existing experimental data and 3--point sum rule (3PSR).

\def\bos{\lower 0.4cm\hbox{{\vrule width 0pt height 1cm}}}
\begin{table}[tbh]
\begin{center}
\begin{tabular}{|c|c|c|}
\hline
                          &\bos $r_1$                        
& $r_2$                       \\ \hline\hline
       E791               &\bos $2.27 \pm 0.35 \pm 0.22$     
& $1.57 \pm 0.25 \pm 0.19$    \\ \hline
       CLEO               &\bos $0.9 \pm 0.6 \pm 0.3$        
& $1.4 \pm 0.5 \pm  0.3$      \\ \hline
       E687               &\bos $1.8 \pm 0.9 \pm0.2$         
& $1.1 \pm 0.8 \pm 0.1$       \\ \hline
       E653               &\bos $2.3^{+1.1}_{-0.9} \pm 0.4$  
& $2.1^{+0.6}_{-0.5} \pm 0.2$ \\ \hline
       Average            &\bos $1.92 \pm 0.32$              
& $1.60 \pm 0.24$             \\ \hline
       3PSR \cite{R6215}  &\bos $2.20 \pm 0.85$              
& $1.16 \pm 0.46$             \\ \hline 
       3PSR \cite{R6216}  &\bos $2.16 \pm  0.38$      
& $-1.08 \pm 0.17$      \\ \hline
       Our Results        &\bos $1.38 \pm 0.44$       
& $1.31 \pm 0.43$      \\ \hline
\end{tabular}
\end{center}
\caption {Comparison of our results for $r_1$ and $r_2$ with the
experimental results and 3--point sum rule.}
\end{table}

Using the parametrization of the $D_s \rar \phi$ transition in terms of the
form factors $A_1$, $A_2$, $V$, $A_3-A_0$, the differential decay width as a
function of $q^2$, in terms of the helicity amplitudes can be written as
\bea
\label{e6226}
\frac{d\Gamma}{dq^2} \es \frac{G_F^2 \vel V_{cs} \ver^2}{192 \pi^3
m_{D_s}^3} \lambda^{1/2}(m_{D_s}^2,m_\phi^2,q^2) q^2
\Big[H_0^2 + H_+^2 + H_-^2 \Big] \nnb \\
\eqv \frac{d\Gamma_L}{dq^2} + \frac{d\Gamma_+}{dq^2}+
\frac{d\Gamma_-}{dq^2}~,
\eea
where the indices in $d\Gamma_i/dq^2$ and $H_i$ denote the polarization of the
$\phi$ meson, $ \lambda(m_{D_s}^2,m_\phi^2,q^2)=(m_{D_s}^2+m_\phi^2-q^2)^2 -
4 m_{D_s}^2 m_\phi^2$, and
\bea
\label{e6227}
H_\pm \es (m_{D_s}+m_\phi) A_1(q^2) \mp
\frac{\lambda^{1/2}(m_{D_s}^2,m_\phi^2,q^2)}{m_{D_s}+m_\phi} V(q^2)~, \\ \nnb \\ 
\label{e6228}
H_0 \es \frac{1}{2 m_\phi \sqrt{q^2}} \Bigg[ (m_{D_s}^2-m_\phi^2-q^2)
(m_{D_s}+m_\phi) A_1(q^2) - \frac{\lambda(m_{D_s}^2,m_\phi^2,q^2)}
{m_{D_s}+m_\phi} A_2(q^2) \Bigg]~.
\eea

The differential decay rate when the final state $\phi$ meson is
transversally polarized is determined to be
\bea
\label{e6229}
\frac{d\Gamma_T}{dq^2} = \frac{d\Gamma_+}{dq^2}+
\frac{d\Gamma_-}{dq^2}~.
\eea
 Integrating the differential decay widths over $q^2$ in the region from
$q^2=0$ to $(m_{D_s}-m_\phi)^2$, we obtain
\bea
\Gamma_L \es (1.52_{+0.77}^{-0.61}) \times 10^{-14}~GeV~, \nnb \\
\Gamma_T \es (2.21_{+1.13}^{-0.90}) \times 10^{-14}~GeV~, \nnb
\eea
and for their ratio, we get
\bea
\frac{\Gamma_L}{\Gamma_T} = (0.69 \pm 0.44)~,\nnb
\eea
which is in good agreement with the existing experimental data 
\bea
\Bigg(\frac{\Gamma_L}{\Gamma_T}\Bigg)_{exp} = 0.72 \pm 0.16~,~\cite{R6221}\nnb
\eea
 
Using the value of the total decay width $\Gamma_{D_s} = 1.34 \times
10^{-12}~GeV$ \cite{R6221} of the $D_s$ meson, we get the following result
for the branching ratio of the $D_s \rar \phi \bar{\ell} \nu$ decay
\bea
{\cal B} (D_s \rar \phi \bar{\ell} \nu) = (2.78_{+1.42}^{-1.13})\%~, \nnb
\eea
which is consistent with the experimental result 
\bea
{\cal B} (D_s \rar \phi \bar{\ell} \nu)_{exp} = (2.0 \pm 0.5)\%~. \nnb 
\eea
  
In conclusion, we calculate the form factors for the $D_s \rar \phi$
transition, in framework of the light cone QCD sum rules. We compare our
results for the form factors with the existing calculations based on
3--point sum rules. Following this analysis, we then estimate the ratios of
these form factors and compare them with the current experimental data, as
well as with the existing theoretical calculations. Finally, we study the
ratio $\Gamma_L/\Gamma_T$ of the decay widths when $\phi$ meson is 
longitudinally and transversally polarized, and the branching ratio. 
Our calculations on the above--mentioned quantities confirm that 
they are consistent with the existing experimental data.

\end{document}